\documentclass[prl,aps,twocolumn,amsfonts,showpacs,superscriptaddress]{revtex4-1}

\newsavebox{\foobox}
\newcommand{\slantbox}[2][0]{\mbox{%
        \sbox{\foobox}{#2}%
        \hskip\wd\foobox
        \pdfsave
        \pdfsetmatrix{1 0 #1 1}%
        \llap{\usebox{\foobox}}%
        \pdfrestore
}}
\newcommand\unslant[2][-.25]{\slantbox[#1]{$#2$}}

\newcommand{\mpi}{\text{\unslant[-.18]\pi}}
\newcommand{\mdelta}{\text{\unslant[-.18]\delta}}

\usepackage{graphicx, isomath}
\usepackage{xcolor}
\usepackage{rotating}
\usepackage{amsmath,amssymb,graphics}

\usepackage[colorlinks=true, urlcolor=violet, linkcolor=blue, citecolor=red, hyperindex=true, linktocpage=true]{hyperref}

\newcommand{\nn}{\nonumber\\}

\def\CO{\mathcal{O}}

\begin{document}
\title{Incoherent thermal transport from dirty black holes}

\author{Sa\v{s}o Grozdanov}
\email{grozdanov@lorentz.leidenuniv.nl}
\affiliation{Instituut-Lorentz for Theoretical Physics, Leiden University, Niels Bohrweg 2, Leiden 2333 CA, The Netherlands}

\author{Andrew Lucas}
\email{lucas@fas.harvard.edu}

\affiliation{Department of Physics, Harvard University, Cambridge MA 02138, USA}

\author{Koenraad Schalm}
\email{kschalm@lorentz.leidenuniv.nl}
\affiliation{Instituut-Lorentz for Theoretical Physics, Leiden University, Niels Bohrweg 2, Leiden 2333 CA, The Netherlands}

\date{\today}

\begin{abstract}
We study thermal transport in strongly disordered, strongly interacting quantum field theories without quasiparticles using gauge-gravity duality.   We analyze linear perturbations of black holes with broken translational symmetry in Einstein-Maxwell-dilaton theories of gravity. Using general geometric arguments in the bulk, we derive bounds on thermal conductivity for the dual disordered field theories in one and two spatial dimensions.   In the latter case, the thermal conductivity is always non-zero at finite temperature, so long as the dilaton potential is bounded from below.   Hence, generic holographic models make  non-trivial predictions about the thermal conductivity in a strongly disordered, strongly coupled  metal in two spatial dimensions.
\end{abstract}

\maketitle

{\it Introduction.---} One of the most dramatic and unique consequences of quantum mechanics is the localization of electronic wave functions in disordered systems \cite{anderson}, which leads to exponentially suppressed thermal (and electrical) conductivity in non-interacting theories at finite temperature \cite{halperin2}.  In recent years, there has been an intense effort to study whether localization extends to interacting theories \cite{basko}.   Under many circumstances, at least in one spatial dimension \cite{huse1, huse2, huse3}, this has been shown to be the case, and the resulting phase is coined many-body localization.  It is of great interest to understand whether this phenomenon persists  in strongly interacting quantum systems in higher dimensions.   

One of the only analytical techniques to study higher dimensional strongly coupled systems is gauge-gravity duality \cite{review1, review2, review3}.  In particular, finite temperature computations become tractable.   In the past few years, numerous holographic models with heuristic ``mean field" approximations to disorder have conjectured (perhaps inadvertently) that the simplest holographic models are thermal conductors \emph{at all disorder strengths}, in two or more spatial dimensions \cite{thermoel1, thermoel3, thermoel2, gouteraux2}.  Such models correctly approximate weak disorder \cite{btv, lss, lucas}, as the theories are in a hydrodynamic regime.   Numerical work on holographic thermal transport \cite{donos1409, rangamani, santosdisorder3} has also been performed by explicitly constructing disordered black holes \cite{santosdisorder1, santosdisorder2}.  What remains an open question is whether mean field results are also sensible when disorder is strong.    

Recently, the tools to analytically address this problem have been developed.   Firstly, the computation of thermal DC conductivity has been reduced to a hydrodynamics problem for an ``artificial" fluid on the black hole horizon \cite{donos1506, donos1507}.  Secondly, a general hydrodynamic framework has been developed which can provide non-perturbative bounds on transport coefficients \cite{lucas4}.  Recently, these formalisms have been combined to prove that the DC electrical conductivity of Einstein-Maxwell holographic models in two spatial (boundary) dimensions is strictly finite \cite{grozdanov}, so long as the black hole horizon is connected.   The goal of this letter is to obtain similar results for the thermal conductivity of disordered Einstein-Maxwell-dilaton (EMD) holographic models.

{\it Results.---} 
We bound the thermal conductivity $\kappa$ of a relativistic (at high energies) theory dual to a disordered black hole in EMD gravity in $d+2$ spacetime dimensions, for $d=1,2$.   These holographic models are dual to field theories at finite density and temperature, deformed by a charge-neutral relevant scalar operator.   The dual theory is sourced by arbitrarily strong disorder, so long as in the bulk, the horizon remains connected. Assuming isotropy on long length scales, $\kappa$ is defined as 
\begin{equation}
\kappa \equiv \left.- \frac{Q_x}{ \partial_x T}\right|_{J_x=0},   \label{kappaeq}
\end{equation}
where $Q_x$ is the heat current, $J_x$ is the charge current, and $-\partial_x T$ is an externally imposed uniform temperature gradient.  Note that $\kappa$ is positive by the second law of thermodynamics, has (relativistic) mass dimension $[\kappa]=[\mathrm{M}]^{d-1}$, and that it is defined under the boundary condition that no average electric current flows.  The boundary theory is at a uniform temperature $T$.

Analytically computing bounds on $\kappa$,  in $d=1$ we find 
\begin{equation}
\kappa  \geq 16\mpi^3 \left( \frac{1}{1-\frac{1}{2}V_{\mathrm{min}}} \right)\frac{T}{s} \times \frac{k_{\mathrm{B}}^2}{\hbar} \left(N^2\right)^2 ,  \label{k1}
\end{equation}
where $s$ is the (spatially averaged) entropy density in the boundary theory, $N^2$ is a large prefactor due to the many degrees of freedom in holographic models, and $V_{\mathrm{min}}$ denotes the global minimum of the dilaton potential; note that $V_{\mathrm{min}} \le 0$ by construction. In $d=2$ we find  
\begin{equation}
\frac{\kappa}{T} \ge \frac{4\mpi^2 }{3} \left( \frac{1}{1- \frac{1}{6} V_{\mathrm{min}}} \right)\times \frac{k_{\mathrm{B}}^2}{\hbar} N^2 .  \label{k2f}
\end{equation}
Hence, if the dilaton potential is bounded from below, the (relativistically) dimensionless number $\kappa/T$ is strictly finite. Such sharp exact results for $\kappa$ have never before been obtained in higher dimensions in a theory where (generic) disorder and interactions are both treated non-perturbatively at finite temperature and density.   Henceforth, we work in units where $\hbar=k_{\mathrm{B}}=c=1$.

Let us briefly compare (\ref{k2f}) with our previous result \cite{grozdanov} showing that in $d=2$, $\sigma$ is strictly finite in Einstein-Maxwell models.  The bound on $\sigma$ derived in \cite{grozdanov} relies entirely on curious properties of holographic ``horizon fluids" governing transport, but not on details of the near-horizon geometry!  Crudely speaking, the horizon fluid is always a local conductor, and so is necessarily a global conductor, as is the dual theory.  No such local bound exists for $\kappa$;  the bulk Einstein's equations play an essential role in enforcing the bound \eqref{k2f}.  For this reason, it is much more subtle.  It is also more powerful:  while in general EMD models $\sigma(T) \sim T^p$ can be either metallic ($p<0$) or insulating ($p>0$)  \cite{rangamani, donos2, gouteraux, donos1406},   we show that many EMD models are never better thermal insulators than naive dimensional analysis predicts.   Many models with IR hyperscaling-violating exponent $\theta\ge 0$ have $V_{\mathrm{min}} > -\infty$ \cite{lucas1411}; for these models \eqref{k2f} is non-trivial.

Our result \eqref{k2f} in $d=2$ is reminiscent of the proposal for ``incoherent" metallic transport presented in \cite{hartnoll1}, whereby a metal with badly broken translational invariance may remain a conductor of thermal (and electric) currents.   The incoherent metal was proposed as a conceptual framework to understand how a strongly interacting system, such as the strange metal phase of the high-$T_{\mathrm{c}}$ superconductors \cite{taillefer}, remains a conductor even when disorder is strong enough to destroy the Drude peak.  In an incoherent metal, transport is diffusion-limited, and remains finite even with strong disorder.   This phenomenon is challenging to realize using traditional condensed matter approaches, as interaction strength and disorder strength cannot both be treated non-perturbatively.    Mean field holographic approaches have recovered incoherent thermal transport before \cite{gouteraux2}.   Our results demonstrate that generic disordered holographic models can realize incoherent thermal transport.

Evidence for hydrodynamic electronic flow in a strongly coupled metal has emerged in experiments on clean graphene \cite{crossno, lucas3}.   However, this hydrodynamic behavior is limited by disorder in real systems, and the possibility for holography to capture transport physics beyond the hydrodynamic limit of \cite{lucas3} is intriguing.   Our qualitative bound $\kappa/T\gtrsim$ constant in $d=2$ may therefore be relevant to ``messier" materials.  It also is interesting to extend these ideas to transport at finite frequency and magnetic field \cite{hkms, lucasMM, magtrans1, magtrans2, magtrans3,  davison15, blake2, blake3, donos1511}.

Returning to our opening question of whether holographic models may realize a many-body localized state,  our work rules this out under a broad set of circumstances:  $\kappa/T$ is bounded, regardless of disorder strength, and not vanishing with increasing disorder.   Indeed, it appears likely that the only possible holographic model of localization contains black holes with horizons of disconnected topologies.   Although some progress has been made in this direction \cite{anninos, santos, qi}, this phenomenon has not been shown to be generic, nor are the properties of such holographic models well understood. Hence, further analyses and constructions of such fragmented black holes remain an interesting and important direction for future work.

{\it Dirty Black Holes.---} We now present the details of our derivation of the bounds \eqref{k1} and \eqref{k2f} on the thermal conductivity $\kappa$. We will consider a family of theories dual to EMD holographic models, for which the Lagrangian is
\begin{equation}\label{ESAct}
\mathcal{L} = R-2\Lambda - \frac{1}{2}(\partial \Phi)^2 - \frac{V(\Phi)}{\ell^2} - \frac{Z(\Phi)F^2}{4e^2},
\end{equation} 
with $\Lambda = -d(d+1)/2 \ell^2$ a negative cosmological constant.  Here $\Phi$ is the scalar dilaton and $F$ is the Maxwell tensor for a U(1) gauge field.  The prefactor $N^2$ defined previously is $N^2 = \ell^2/16\mpi G_{\mathrm{N}}$.  We assume $Z(\Phi=0)=1$ and $V(\Phi = 0) = 0$;  $V_{\mathrm{min}}$ as defined previously is $V_{\mathrm{min}} = \min_\Phi(V(\Phi))$, which is assumed to be finite, i.e. the dilaton potential is everywhere bounded from below.  We have set $16\mpi G_{\mathrm{N}}=1$, and will also set $e=\ell=1$ henceforth \footnote{Note that $\ell^2/16\mpi G_{\mathrm{N}} = N^2$, but since $N^2$ only enters the computation as an overall prefactor in $\kappa$, it is not necessary to keep track of it explicitly.}.

It is most convenient to consider static background black hole geometries in Gaussian normal coordinates: 
\begin{equation}\label{GNmetric}
\mathrm{d}s^2 =  - f(r,\mathbf{x})^2 \mathrm{d}t^2 + \mathrm{d}r^2 + G_{ij}(r,\mathbf{x})\mathrm{d}x^i \mathrm{d}x^j,
\end{equation}
with a connected black hole horizon at $r=0$ \footnote{To obtain the near-horizon Gaussian normal coordinates from the coordinates $\left(t, \tilde r,\tilde{\mathbf{x}}\right)$ that were used in \cite{donos1506, donos1507,grozdanov}, we write $\tilde r = \Gamma_0(\mathbf{x}) r^2 + \Gamma_1(\mathbf{x}) r^3 +\CO\left(r^4\right)$, $\tilde x^i = x^i + \Omega^i_0 (\mathbf{x}) r^2  +\CO\left(r^3\right)$ and appropriately adjust the functions $\Gamma_n(\mathbf{x})$ and $\Omega^i_n(\mathbf{x})$.   Importantly, this process does not alter $\gamma_{ij}$ or $\phi$, nor the leading order behavior in $A_t$ up to a simple rescaling.}. The coordinate $r$ denotes the holographic radial direction ($r$ increases towards  the boundary, which we take to be asymptotically AdS), and $(t,\mathbf{x})$ are boundary theory directions.   The (uniform) temperature of the black hole horizon is $T$, which is also the temperature of the dual field theory.   This follows from the zeroth law of black hole thermodynamics and is true for any amount of disorder. 
Assuming a smooth horizon, we can constrain the near-horizon expansion of $f$ and $G$ (see e.g. \cite{medved}): 
\begin{subequations}\label{nearhor}
\begin{align}
f(r,\mathbf{x}) &= 2\mpi T r + \frac{F(\mathbf{x})}{6}r^3 + \cdots,  \\
G_{ij}(r,\mathbf{x}) &= \gamma_{ij}(\mathbf{x}) + \frac{h_{ij}(\mathbf{x})}{2}r^2 + \cdots .
\end{align}
\end{subequations} 
A $d$-dimensional diffeomorphism symmetry $x_i \rightarrow X_i(x_j)$ is still remaining, and we may use this to our advantage. The dilaton admits the near-horizon expansion 
\begin{equation}\label{Scalarnearhor}
\Phi(r,\mathbf{x}) = \phi(\mathbf{x}) +  \cdots
\end{equation}
and the Maxwell field \begin{equation}
A = \left(\mpi T \mathcal{Q} r^2 + \cdots\right) \mathrm{d}t.
\end{equation}
Although our choice of radial coordinate is a bit different than what is used in \cite{donos1507}, since this metric obeys all general constraints demanded in their paper, we may still use their results in our new coordinate system.  We assume our boundary theory and black hole horizon have topology $\mathrm{T}^d$, and so we can introduce normalized spatial averaging over the horizon, which we denote by 
\begin{equation}
\mathbb{E}[\circ] = \frac{1}{L^d} \int \mathrm{d}^d\mathbf{x} \; \circ.
\end{equation}
In the above definition, we have not included a factor of the induced horizon metric.  This convention follows \cite{grozdanov}.  The boundary spatial torus has a length $L$ in each direction, and spatial metric $\mdelta_{ij}$, for simplicity.

The requirement that the Ricci scalar curvature of the induced horizon metric be non-singular implies that the extrinsic curvature of the horizon must vanish. As a result, the horizon is the hypersurface (at constant $t$) with minimal area $A_{\text{hor}}$ (see e.g. \cite{solodukhin}). In our setup, the black hole horizon is a $d$-dimensional hypersurface of topology $\mathrm{T}^d$ on a fixed time slice.  Let us now consider the hypersurface of points lying along the curve $r=\varepsilon R(\mathbf{x})$, with $R(\mathbf{x})\ge 0$ and $\varepsilon$ an infinitesimally small positive number. This hypersurface thus lies just outside of the horizon for all $\mathbf{x}$ and must as a consequence of our above discussion obey
\begin{align}\label{GeoBound}
A_{\text{hor}} \leq  A_{R} . 
\end{align}
When $\varepsilon$ is small, the area $A_R$ is  
\begin{align}\label{HyperArea}
A_R \approx  \int \mathrm{d}^d\mathbf{x}\sqrt{ \det\left(\gamma_{ij} + \frac{ \varepsilon^2}{2}h_{ij}R^2 + \varepsilon^2 \partial_i R \, \partial_j R \right)}  .
\end{align}
Note that $A_{\mathrm{hor}}$ is obtained by setting $R=0$.   Expanding \eqref{HyperArea} to leading order in $\varepsilon$ and using \eqref{GeoBound} gives  
\begin{equation}
0 \leq \mathbb{E}\left[\sqrt{\gamma}\left(\gamma^{ij} \partial_i R \, \partial_j R + \frac{\gamma^{ij}h_{ij}}{2} R^2\right)\right] , \label{rtbound}
\end{equation}
which should hold for all possible choices of $R(\mathbf{x})\geq 0$. The inequality in Eq. \eqref{rtbound} allows us to place non-trivial constraints on the possible $h_{ij}$, and will play a crucial role in finding bounds on $\kappa$.

{\it One Dimension.---} We begin by considering boundary theories with one spatial dimension, where it has recently been shown \cite{donos1507} that given any EMD theory \eqref{ESAct}:\footnote{In this formula, and in (\ref{donoseq}), we have already transformed the expression to our coordinate system.}
\begin{equation}
\kappa = \frac{16\pi^2 T}{\mathbb{E}\left[\gamma^{-1/2}(\partial_x \phi)^2 + \gamma^{1/2} Z(\phi)\mathcal{Q}^2\right]}.
\end{equation}
In this case, the Gaussian normal coordinates used in \eqref{GNmetric}, \eqref{nearhor} and \eqref{Scalarnearhor} prove to be particularly convenient.  By using the leading order coefficients of the $tt$ component of Einstein's equations near the horizon, we find the following: 
\begin{align}
&\mathbb{E}\left[\frac{(\partial_x \phi)^2}{\sqrt{\gamma}} + \sqrt{\gamma}Z(\phi)\mathcal{Q}^2\right] \nn
&= \mathbb{E}\left[4\sqrt{\gamma} - \frac{2h_{xx}}{\sqrt{\gamma}} - 2\sqrt{\gamma} \,V(\phi) \right].
\end{align}
By substituting the ansatz $R=1$  into the bound \eqref{rtbound}, we find 
\begin{equation}
\mathbb{E}\left[\frac{h_{xx}}{\sqrt{\gamma}}\right] \ge 0.  \label{b1}
\end{equation}
Using $s = 4 \mpi \mathbb{E}[\sqrt{\gamma}]$ and $V\geq V_{\mathrm{min}}$ directly implies (\ref{k1}).   

{\it Two Dimensions.---} Let us now consider the thermal conductivity of a disordered black hole in EMD gravity when $d=2$.   We assume the black hole geometry is isotropic (in the boundary spatial directions) on thermodynamically large length scales.  Following \cite{lucas4, grozdanov}, it is straightforward to derive a variational bound on the thermal conductivity.   \cite{donos1507} showed that the DC transport coefficients may be computed as follows: solve the following equations
\begin{subequations}\label{donoseq}\begin{align}
&\nabla_i \mathcal{I}^i = \nabla_i \mathcal{J}^i = 0, \\
&\mathcal{I}_i = Z(\phi) \left(E_i - \nabla_i \mu\right) + (4\mpi T)^{-1} Z(\phi)\mathcal{Q} \mathcal{J}_i,  \\
&Z(\phi)\mathcal{Q}\left(\nabla_i \mu-E_i \right) + 4\mpi \left( \nabla_i \Theta-T\zeta_i \right)  = \notag \\ 
&~~~~~~~~~~(4\pi T)^{-1}\left[2\nabla^j\nabla_{(j} \mathcal{J}_{i)} - \mathcal{J}^j \nabla_j \phi \nabla_i \phi\right],
\end{align}\end{subequations}
where $\phi$ and $\mathcal{Q}$ are the horizon data defined previously, covariant derivatives are taken with respect to the induced horizon metric $\gamma_{ij}$,  $\mu$ and $\Theta$ are scalar functions and $E_i$ and $\zeta_i$ are an applied electric field and thermal ``drive" ($\zeta_i$ is analogous to $-\partial_i \log T$) in the boundary theory.   The spatially averaged boundary charge and heat currents are 
\begin{align}
J_{\mathrm{charge}}^i = \mathbb{E}[\sqrt{\gamma}\mathcal{I}^i], && J_{\mathrm{heat}}^i = \mathbb{E}[\sqrt{\gamma}\mathcal{J}^i].  \label{jcharge}
\end{align}
The Joule heating in the boundary theory is given by \begin{align}
\mathcal{P} &= E_i J^i_{\mathrm{charge}}+ \zeta_i J^i_{\mathrm{heat}} \notag \\
&= \frac{1}{16\mpi^2T^2}\mathbb{E}\biggr[2\sqrt{\gamma}\nabla^{(i}\mathcal{J}^{j)}\nabla_{(i}\mathcal{J}_{j)} + \sqrt{\gamma} \left(\mathcal{J}^i \nabla_i \phi\right)^2 \notag \\
&+\sqrt{\gamma} Z(\phi)\left(\mathcal{Q}\mathcal{J}_i -\frac{4\mpi T}{Z(\phi)} \mathcal{I}_i\right) \left(\mathcal{Q}\mathcal{J}^i -\frac{4\mpi T}{Z(\phi)}\mathcal{I}^i\right) \biggr],  \label{vareq}
\end{align}
as is simple to check by integrating each term in the above expression by parts, and employing  (\ref{donoseq}) and (\ref{jcharge}).

We now think of $\mathcal{P}$ as a functional of arbitrary currents $\mathcal{I}^i$ and $\mathcal{J}^i$ which are conserved.   If $\overline{\mathcal{I}}$ and $\overline{\mathcal{J}}$ are the currents that solve (\ref{donoseq}) for fixed $J_{\mathrm{charge}}$ and $J_{\mathrm{heat}}$,   then writing an arbitrary current as $\mathcal{I} = \overline{\mathcal{I}} + \tilde{\mathcal{I}}$ and $\mathcal{J} =  \overline{\mathcal{J}} + \tilde{\mathcal{J}}$ we find that (using that $\overline{\mathcal{I}}$ and $\overline{\mathcal{J}}$ obey (\ref{donoseq})) 
\begin{align}
\mathcal{P} \left[\mathcal{I},\mathcal{J}\right] &= \mathcal{P}\left[\overline{\mathcal{I}},\overline{\mathcal{J}}\right] + \mathcal{P}\left[\tilde{\mathcal{I}}, \tilde{\mathcal{J}}\right] \nn
&+ \mathbb{E}\left[2\sqrt{\gamma} \left(E_i \tilde{\mathcal{I}}^i + \zeta_i \tilde{\mathcal{J}}^i\right)\right].
\end{align}
The last term vanishes if we fix  normalization (\ref{jcharge}).   As (\ref{vareq}) is positive definite, we conclude that $\mathcal{P} \ge \mathcal{P}[\overline{\mathcal{I}},\overline{\mathcal{J}}]$.  Furthermore, the power dissipated in the absence of electric current is $J_{\mathrm{heat}}^2/T\kappa$.   Hence, if $\mathcal{I}$ and $\mathcal{J}$ are trial currents which are conserved and normalized to
$J_{\mathrm{heat}}=1$ and $J_{\mathrm{charge}}=0$, we may employ (\ref{vareq}) to find an upper bound on $1/\kappa$.
A simple guess for $\mathcal{I}$ and $\mathcal{J}$ is
\begin{align}
 \mathcal{I}^i = 0,&& \mathcal{J}^i  = \frac{1}{\sqrt{\gamma}}\mdelta^i_x .  \label{varans}
\end{align}

It is helpful to use the residual diffeomorphism invariance to fix \begin{equation}
\gamma_{ij} = \mdelta_{ij} \mathrm{e}^{\omega(\mathbf{x})}.  \label{confmet}
\end{equation}   Because of the emergent isotropy of our model, this diffeomorphism should not change the length of the spatial torus in either direction in the thermodynamic limit (we may add a constant to $\omega$ to achieve this). Combining our variational ansatz (\ref{varans}) with the induced horizon metric choice (\ref{confmet}), we obtain the bound 
\begin{align}
\frac{T}{\kappa} \le&\, \frac{1}{16\mpi^2 }\,\mathbb{E}\left[\mathrm{e}^{-\omega} \left((\partial_x \omega)^2 + (\partial_y\omega)^2 + (\partial_x \phi)^2 \right) + Z(\phi) \mathcal{Q}^2\right] \nn
=&\, \frac{1}{16\mpi^2 }\,\mathbb{E}\left[- R_2 + \mathrm{e}^{-\omega}(\partial_x\phi)^2 + Z(\phi)\mathcal{Q}^2\right],   \label{k2}
\end{align}
where $R_2$ is the induced Ricci scalar on the horizon:
\begin{align}\label{R2exp}
R_2 = -\mathrm{e}^{-\omega} \left (\partial_x^2 \omega + \partial_y^2\omega\right).
\end{align} 

We now derive an upper bound on $\mathbb{E}\left[R_2\right]$. The four dimensional Ricci scalar $R_4$ in the bulk is given by
\begin{equation}\label{B41}
R_4 = - 12 + \frac{1}{2}(\partial \Phi)^2 + 2V(\Phi),
\end{equation}
for any solution of the Einstein's equations that follow from the Lagrangian \eqref{ESAct}. General arguments \cite{medved} about geometries near black hole horizons, independent of the equations of motion, then allow us to relate $R_4$ to $R_2$. At the horizon, i.e. at $r=0$,
\begin{equation}\label{B42}
R_4 = R_2 - \frac{F}{\mpi T} - 2\mathrm{e}^{-\omega}(h_{xx}+h_{yy}).
\end{equation}
Einstein's equations at leading order in the near-horizon expansion then force 
\begin{equation}\label{B43}
\frac{F}{\mpi T} = 6 + \frac{Z(\phi)\mathcal{Q}^2}{2} - \mathrm{e}^{-\omega}(h_{xx}+h_{yy}) - V(\phi).
\end{equation}
Combining  \eqref{B41}, \eqref{B42} and \eqref{B43} at $r=0$, we obtain 
\begin{align}
R_2 &= -6 +  \mathrm{e}^{-\omega}(h_{xx}+h_{yy}) \nn
&+ \frac{Z(\phi)\mathcal{Q}^2}{2} + \frac{(\partial \phi)^2}{2} + V(\phi).  \label{r26}
\end{align}
Next, we plug the  ansatz $R(\mathbf{x})=\mathrm{e}^{-\omega(\mathbf{x})/2}$ into (\ref{rtbound}), which gives us the inequality 
\begin{align}\label{HyperIneq2}
0 \le \mathbb{E} & \left[ \mathrm{e}^{-\omega} \left(  \frac{1}{2}\left( \partial_x \omega \right)^2 + \frac{1}{2} \left( \partial_y \omega \right)^2  +h_{xx} + h_{yy} \right) \right].
\end{align} 
Recognizing $R_2$ from Eq. \eqref{R2exp} after partial integration inside \eqref{HyperIneq2}, we find
\begin{equation}
\mathbb{E}\left[R_2\right]  \le 2 \,\mathbb{E}\left[\mathrm{e}^{-\omega} (h_{xx} + h_{yy})\right] .   \label{b2}
\end{equation}
Combining  (\ref{r26}) and (\ref{b2}), along with $V\ge V_{\mathrm{min}}$, we obtain 
\begin{align}
&\mathbb{E}\left[  \mathrm{e}^{-\omega}(\partial_x \phi)^2 + Z(\phi)\mathcal{Q}^2-R_2\right] \notag \\
&\leq \mathbb{E}\left[ (\partial \phi)^2 + Z(\phi)\mathcal{Q}^2 - R_2\right] \leq 12-2V_{\mathrm{min}},
\end{align}
which can be used in  \eqref{k2} to establish   (\ref{k2f}).  In both $d=1$ and $d=2$, we may replace $V_{\mathrm{min}}$ by $\mathbb{E}[V(\phi)]$, if we are able to compute this quantity.    In particular, even if $V(\Phi \rightarrow \pm \infty)$ is unbounded,  $\kappa$ is finite so long as $\phi$ is finite everywhere at finite $T$.   There can be no analogue of many-body localization if $V(\phi)$ is finite on the horizon.

A simple extension of the model we have studied so far in this paper includes $n>1$ scalar fields $\Phi^a$ in $d=2$, 
\begin{equation}\label{ESAct2}
\mathcal{L} =  R-2\Lambda -  \sum_{a=1}^n \frac{\left(\partial \Phi^a\right)^2}{2} - V(\Phi^a)- \frac{Z(\Phi^a)F^2}{4}.
\end{equation} 
The bounds for the theory in \eqref{ESAct2} are still \eqref{k1} and \eqref{k2f}. 

{\it Mean Field Models.---} 
Let us now compare our exact results with the predictions of ``mean field" models of disorder in holography. A simple model in $d=2$ consists of $n=2$ copies of the same scalar field, with Lagrangian (\ref{ESAct2}), but with $V=0$ and $Z=1$ \cite{andrade, donos1}.  Choosing the scalar fields to be $\Phi^i = m x^i$,  one finds \cite{thermoel3}
\begin{equation}
\kappa = \frac{4\mpi s T}{m^2+\mu^2},
\end{equation}
where $s$ is the entropy density and $\mu$ is the chemical potential. In order for the temperature of the dual theory to be non-negative, \begin{equation}
 2m^2 + \mu^2 \leq \frac{3}{\mpi} s.
\end{equation}
As it must, our bound \eqref{k2f} is obeyed, as \begin{equation}
\kappa \ge \frac{4\mpi^2 T}{3} \frac{2m^2+\mu^2}{m^2+\mu^2} \ge \frac{4\mpi^2 T}{3}
\end{equation}
The fact that mean field models capture the correct behavior of $\kappa$ in explicitly disordered models is a pleasing surprise to us.

{\it Acknowledgements.---} We thank Jerome Gauntlett, Blaise Gout\'eraux, Sean Hartnoll, Moshe Rozali and Subir Sachdev for helpful discussions.   AL is supported by the NSF under Grant DMR-1360789 and MURI grant W911NF-14-1-0003 from ARO.   SG and KS are supported by a VICI grant of the Netherlands Organization for Scientific Research (NWO), by the Netherlands Organization for Scientific Research/Ministry of Science and Education (NWO/OCW) and by the Foundation for Research into Fundamental Matter (FOM).    

\bibliographystyle{unsrt}
\bibliography{disorderbib2}
\end{document}